\documentstyle[12pt]{article}

\begin{document}

\begin{titlepage}
\begin{center}

\bigskip
\vspace{3\baselineskip}

{\Large \bf

Spontaneous reduction of noncommutative gauge symmetry and model
building \\}

\bigskip

\bigskip

{\bf Masud Chaichian$^{\mathrm{a}}$,  Archil Kobakhidze$^{\mathrm{a,b}}$
and Anca Tureanu$^{\mathrm{a}}$ \\}
\smallskip

{ \small \it
$^{\mathrm{a}}$High Energy Physics Division, Department of Physical Sciences,
University of Helsinki $\&$ \\
Helsinki Institute of Physics, FIN-00014 Helsinki, Finland\\
$^{\mathrm{b}}$Andronikashvili Institute of Physics, Georgian Academy of Sciences,
GE-380077 Tbilisi, Georgia\\}

\bigskip

\vspace*{.5cm}

{\bf Abstract}\\
\end{center}
\noindent We propose a mechanism for the spontaneous
(gauge-invariant) reduction of noncommutative ${\cal U}(n)$ gauge
theories down to $SU(n)$. This can be achieved through the
condensation of composite ${\cal U}(n)$ gauge invariant fields
that involves half-infinite Wilson lines in trace-$U(1)$
noninvariant and $SU(n)$ preserving direction. Based on this
mechanism we discuss anomaly-free fully gauge invariant
noncommutative Standard Model based on the minimal gauge group
${\cal U}(3)\times {\cal U}(2)\times {\cal U}(1)$, previously
proposed, and show how it can be consistently reduced to the
Standard Model with the usual particle spectrum. Charge
quantization for quarks and leptons naturally follows from the
model.

\bigskip


\bigskip

\end{titlepage}

\subparagraph{Introduction.}

Noncommutative (NC) space naturally emerges in string theory in
the presence of non-zero background B-field (see e.g. the reviews
\cite {Seiberg:1999vs, Douglas:2001ba, Szabo:2001kg} and
references therein). If we seriously accept this possibility an
important task is to 'reproduce' the known physics at low energies
which is described by the celebrated Standard Model with an
amusing accuracy. The construction of a consistent noncommutative
Standard Model (NCSM) however, faces significant difficulties. One
is related with restrictions imposed by noncommutative group
theory and gauge invariance. Namely : (i) Only ${\cal
U}(n)$\footnote{Calligraphic letters denote noncommutative gauge
groups (e.g. ${\cal U}(n)$), while italic letters denote
commutative groups (e.g. $SU(n)$).}
unitary gauge theories (as well as direct product of different ${\cal U}%
(n_{i})^{\prime }$s, $\prod_{i=1}^{k}\times {\cal U}(n_{i})$)
admit noncommutative extension \cite{Matsubara:2000gr}\footnote{%
Recall that in noncommutative case ${\cal U}(n)$ $\neq SU(n)\times
U(1)$, while $U(n)=SU(n)\times U(1)$ in the commutative case.},
but not $SU(n)^{\prime }$s; (ii) Non-trivial representations of
noncommutative ${\cal U}(n)$ are constrained to be fundamental
(left module), antifundamental (right module) or adjoint
(left-right module) only. In
addition, the only allowed non-trivial representations of the product of gauge groups $%
\prod_{i=1}^{k}\times {\cal U}(n_{i})$ are those transforming as
fundamental - antifundamental under the two ${\cal U}(n_{i})$
factors at most \cite{ Terashima:2000xq,Chaichian:2001mu,
Chaichian:2001py}.

An interesting way of circumventing these group-theoretical
problems is through an alternative approach to NC gauge theories
based on the expansion in NC parameter and Seiberg-Witten map.
This approach admits NC $SU(n)$ gauge theories
\cite{Jurco:2001rq}. The model building along these alternative
approach can be found e.g. in \cite {Calmet:2001na,
Aschieri:2002mc}.

However, just from the above group-theoretic properties it is
evident that straightforward (based on Weyl-Moyal approach)
noncommutative extension of
the Standard Model gauge group (that is, $G_{NCSM}={\cal U}(3)\times {\cal U}%
(2)\times {\cal U}(1)$ ) already contains new particles -- two extra
gauge bosons (the rank of $G_{NCSM}$ is 6 vs 4 of
$G_{SM}=SU(3)\times SU(2)\times U(1)$ ). Beside that, there is a
problem of matter (quark-lepton) representations as
well. Namely, since the only allowed charges within the noncommutative $%
{\cal U}(1)$ are 0,$\pm 1$ \cite{Hayakawa:1999yt}, it is clear
that ${\cal U}(1)$ can not be identified with usual weak
hypercharge to account for the fractional electric
charges of the quarks. Hence a different embedding of the electric charge in $%
G_{NCSM}$ must be found.

An attempt to solve these problems has been made in
\cite{Chaichian:2001py}. The extra gauge bosons are made massive,
leaving at low energies just SM gauge group $G_{SM}.$ This was
achieved by introduction of the so-called Higgsac fields which
transform under the trace-$U(1)$ parts of $G_{NCSM}$. The matter
content has been chosen exactly as in the usual Standard Model, but
now obeying the no-go theorem \cite{Chaichian:2001mu}\footnote{As
far as the particle classification is concerned, the use of the
representations of the usual Poincar\'e symmetry has been recently
justified, when it was noticed that noncommutative field theories
with commutation relation $[x_\mu,x_\nu]=i\theta_{\mu\nu}$, with
$\theta_{\mu\nu}$ an antisymmetric constant matrix, are invariant
under twisted Poincar\'e algebra \cite{twist}, whose representations
are the same as those of the usual Poincar\'e group.}. Remarkably,
upon the Higgsac condensation a linear combination of trace-$U(1)
$'s in $G_{NCSM}$ which remains massless is just the weak
hypercharge and thus the fractional charges of quarks are explained
automatically. This is a very welcome outcome of the model and
somehow reminds the charge
quantization within the usual commutative models of grand unification%
\footnote{%
Another approach to the charge quantization problem is to find a
different embedding of the electric charge in an extended
noncommutative gauge symmetry. This has been recently discussed in
\cite{Khoze:2004zc} within a model with ${\cal U}(4)\times {\cal
U}(3)\times {\cal U}(2)$ gauge symmetry, where several conditions
are fulfilled. Besides the extended gauge group, however, the model
requires the introduction of 3 extra generations of mirror quarks
and leptons in order to achieve anomaly cancelation.}.

Unfortunately, the above nice picture has a serious theoretical
drawback. The point is that the symmetry breaking by Higgsac field
is not spontaneous. As a result, one accounts for the violation of
unitarity in gauge boson scattering at high energies
\cite{Hewett:2001im}. Another problem of the model of ref.
\cite{Chaichian:2001py} is that it contains gauge anomalies
related with extra trace-$U(1)$'s in $G_{NCSM}.$ As usually, one
can add extra matter fields to make each $G_{NCSM}$ factor
vector-like and hence the whole theory anomaly-free. Upon the
symmetry breaking down to the $G_{SM}$ these extra matter is
vector-like, and, in principle, can pick up mass through the
Yukawa couplings with the appropriate Higgsac fields. But once
again these Yukawa couplings are not $G_{NCSM}$ gauge invariant.
Summarizing the above discussion, it seems that the problems of
the model of ref. \cite {Chaichian:2001py} can be avoided by
finding a proper gauge invariant realization of the Higgsac
mechanism. Below we discuss such a mechanism involving NC Wilson
lines.

\subparagraph{Spontaneous NC gauge symmetry breaking.}

Consider 'canonical' NC space-time which is defined through the $*-$%
commutation relations,
\begin{equation}
\left[ x_{\mu },x_{\nu }\right] _{*}=i\theta _{\mu \nu }\ ,
\label{1}
\end{equation}
where $\theta_{\mu\nu}$ is an antisymmetric constant matrix.
The $x_{\mu }$ in (\ref{1}) are the ordinary 4-coordinates with $*-$%
multiplication defined as:
\[
f(x)*g(x)=\left. \exp \left( \frac{i}{2}\theta _{\mu \nu }\partial _{x_{\mu
}}\partial _{y_{\nu }}\right) f(x)g(y)\right| _{x=y}.
\]
On this NC space-time we define NC gauge theory based on the gauge group $%
{\cal U}(n)$. The $n^{2}$ gauge bosons form an adjoint representation of $%
{\cal U}(n)$ :
\begin{equation}
A_{X}^{\mu }(x)\longrightarrow u(x)*\left( A_{\mu }(x)-\frac{i}{g}{\bf 1}%
_{n\times n}\partial _{\mu }\right) *u^{-1}(x),  \label{2}
\end{equation}
where $u(x)=\exp _{*}(-ig\beta ^{A}(x)T^{A})$ is an element (defining
representation) of ${\cal U}(n)$ group, $A_{\mu }(x)=A_{\mu }^{A}(x)T^{A}$
is an $u(n)$-algebra valued gauge field with generators $T^{A}=\frac{1}{2}%
\lambda ^{A}$, where $\lambda ^{1},...,\lambda ^{n^{2}-1}$ are the
generalized Gell-Mann matrices and $T^{0}={\bf 1}_{n\times n}$, and
$g$ is the gauge coupling constant.

Recall that commutative $U(n)$ gauge symmetry is
broken spontaneously down to the $SU(n)$ subgroup once a $SU(n)$-singlet and $%
U(1)$-charged scalar field acquires non-zero vacuum expectation
value. One of such allowed (in commutative case) representations is
$n$-index totally antisymmetric tensor
representation\footnote{Notice that, due to the constraints on the
representations of NC groups
\cite{Terashima:2000xq,Chaichian:2001mu} (i.e. matter fields can be
only in fundamental, antifundamental, adjoint or singlet
representations of NC ${\cal U}(n)$), the auxiliary tensor
representation in (\ref{3}) is not a representation of the NC ${\cal
U}(n)$ gauge group.}
\begin{equation}
\phi ^{\left[ i_{1}i_{2}...i_{n}\right] }(x)\ ,  \label{3}
\end{equation}
out of which the scalar field $\phi(x)$ can be constructed in the
form
\begin{equation}
\phi (x)=\frac{1}{n!}\epsilon _{i_{1}i_{2}...i_{n}}\phi ^{\left[
i_{1}i_{2}...i_{n}\right] }(x)\ .  \label{3'}
\end{equation}
The field $\phi(x)$ in (\ref{3'}) carries $U(1)$ charge equal to $n$
and is the representation of the Higssac field used in
\cite{Chaichian:2001py}. However, the noncommutative ${\cal
U}(n)$-transformations do not close when acting on the Higgsac field
$\Phi$, and hence the field $\Phi$ is not a representation of the
${\cal U}(n)$ group. Subsequently, the symmetry breaking in
\cite{Chaichian:2001py} is not spontaneous, since it goes through a
gauge non-invariant mechanism.
 To restore the gauge invariance, instead
of (\ref{3'}) we introduce the following scalar field:
\begin{equation}
\Phi (x)=\frac{1}{n!}\epsilon
_{i_{1}i_{2}...i_{n}}W_{j_{1}}^{i_{i}}*W_{j_{2}}^{i_{2}}*...*W_{j_{n}}^{i_{n}}*\phi
^{\left[ j_{1}j_{2}...j_{n}\right] }(x)\ , \label{4}
\end{equation}
where
\begin{eqnarray}
W &=&P_{*}\exp \left( ig\int_{0}^{1}d\sigma \frac{d\xi ^{\mu }}{d\sigma }%
A_{\mu }(x+\xi (\sigma ))\right)   \label{5} \\
&=&{\bf 1}_{n\times n}+\sum_{n=1}^{\infty }\frac{\left( ig\right) ^{n}}{n!}%
\int_{0}^{1}d\sigma _{1}\int_{\sigma _{1}}^{1}d\sigma _{2}...\int_{\sigma
_{n-1}}^{1}d\sigma _{n}\frac{\partial \xi ^{\mu _{1}}}{\partial \sigma _{1}}%
...\frac{\partial \xi ^{\mu _{n}}}{\partial \sigma _{n}}A_{\mu _{1}}(x+\xi
(\sigma _{1}))*...*A_{\mu _{n}}(x+\xi (\sigma _{n}))  \nonumber
\end{eqnarray}
is a half-infinite Wilson line, with path ordering defined with
respect to $* $-product, the contour $C$ is:
\[
C=\left\{ \left. \xi ^{\mu }(\sigma ),\ \ \ 0<\sigma <1\right| \xi
^{\mu }(0)=\infty ,\ \ \ \xi ^{\mu }(1)=0\right\} ,
\]
and $\phi ^{\left[ j_{1}j_{2},...,j_{n}\right] }(x)$ is an
antisymmetric $n$-index object under ${\cal U}(n)$. The actual shape
of the Wilson line (\ref{5}) is not important and thus it can be
arbitrary.
Within the physically admissible gauge transformations (i.e. those for which $%
u(x)\rightarrow {\bf 1}$ when $x\rightarrow \infty $) this Wilson
line transforms as an antifundamental object
\begin{equation}
W(x)\rightarrow W(x)*u^{-1}(x)\ .  \label{5b}
\end{equation}
Then the composite field $\Phi$ in (\ref{4}) is a gauge-invariant
object \cite {Chu:2001if, Chu:2001kq}. Using the Taylor expansion
(\ref{5}) of the Wilson lines in (\ref{4}),
\[
\Phi (x)=\phi (x)+....,
\]
we see that the first term in the expansion is just the ordinary
Higgsac field (\ref{3'}), while the rest of the terms provide a
gauge invariant completion. Now, if the field $\Phi (x)$ develops a
non-zero vacuum expectation value along the Higgsac direction, i.e.,
\[
<\Phi (x)>=<\phi (x)>=const.,
\]
the NC ${\cal U}(n)$ gauge symmetry becomes spontaneously broken down to $%
SU(n)$. Indeed, since $\Phi (x)$ is the gauge-singlet field we can write a
simple Lagrangian for it:
\begin{equation}
{\cal L}_{Higgsac}=\partial _{\mu }\Phi \partial ^{\mu }\Phi
^{*}-V(\Phi \Phi ^{*}),  \label{6}
\end{equation}
where
\begin{equation}
V(\Phi \Phi ^{*})=m^{2}\Phi \Phi ^{*}+\frac{\lambda }{2}\left( \Phi *\Phi
^{*}\right) ^{2}  \label{7}
\end{equation}
is the bounded from below ($\lambda >0$) tachyonic potential
($m^{2}<0$). The Lagrangian (\ref{6}) can be viewed as a
gauge-invariant version of the Higgsac Lagrangian proposed in
\cite{Chaichian:2001py}. As in the ordinary commutative case, we
assume that the perturbative vacuum for the gauge field is given
by the pure gauge configuration equivalent to the trivial vector
potential, i.e., $<A_{\mu }>=0$. Then $<W>={\bf 1}_{n\times n}$,
and
the potential (\ref{7}) is reduced to the potential for the Higgsac field $%
\phi (x),$ $V(\phi \phi ^{*})$, with nontrivial minimum that can
be chosen as:
\begin{equation}
<\phi (x)>=\sqrt{-\frac{m^{2}}{\lambda }}.  \label{8}
\end{equation}
Hence we expect that trace-$U(1)$ field of NC ${\cal U}(n)$ gauge theory
picks up a mass leaving $SU(n)$ unbroken. To see this we must closely
inspect the kinetic term in (\ref{6}). First note that the leading order
term ($\theta$-independent) in $\theta$-expansion for
the composite object (\ref{4}) looks as
\[
\Phi (x)=(\det W)\ \phi (x)=\left( 1+ig\int_{0}^{1}d\sigma \frac{d\xi ^{\mu }}{%
d\sigma }TrA_{\mu }(x+\xi (\sigma ))+...\right) \phi (x).
\]
Hence, the expansion of $\partial _{\mu }\Phi (x)$ contains the ordinary
covariant derivative for the Higgsac field, i.e.,
\begin{eqnarray*}
\partial _{\mu }\Phi (x) &=&\left( \partial _{\mu }+ingA_{\mu }^{0}\right)
\phi (x) \\
&&+ig\left[ \int_{0}^{1}d\sigma \frac{d\xi ^{\mu }}{d\sigma }TrA_{\mu
}(x+\xi (\sigma ))\right] \partial _{\mu }\phi (x)+...,
\end{eqnarray*}
along with other terms which again provide the gauge-invariant
completion. Evaluating at the minimum (\ref{8}) the kinetic term
in (\ref{6}) we obtain
the mass for the trace-$U(1)$ gauge boson $A_{\mu }^{0}$, $M_{A^{0}}^{2}=-2%
\frac{n^{2}g^{2}m^{2}}{\lambda }$. This is how the spontaneous symmetry
breaking ${\cal U}(n)\rightarrow SU(n)$ occurs. This can be
straightforwardly generalized to the breaking ${\cal U}(n)\times {\cal U}%
(m)\rightarrow SU(n)\times SU(m)$. In this case we need a
composite Higgsac field which carries charge $n$ coupled to
trace-$U(1)$ of ${\cal U}(n)$ and charge $-m$ coupled to
trace-$U(1)$ of ${\cal U}(m)$, i.e.,
\begin{eqnarray}
\Phi (x)_{{\cal U}(n)\times {\cal U}(m)} &=&\frac{1}{n!m!}\epsilon
_{i_{1}i_{2},...,i_{n}}\epsilon ^{l_{1}l_{2},...,l_{m}}\left( W_{{\cal U}%
(n)}\right) _{j_{1}}^{i_{1}}*\left( W_{{\cal U}(n)}\right)
_{j_{2}}^{i_{2}}*...*\left( W_{{\cal U}(n)}\right) _{j_{n}}^{i_{n}}
\nonumber \\
&&*\phi (x)^{[j_{1}j_{2}...j_{n}]}_{[k_{1}k_{2}...k_{m}]}
*\left( W_{{\cal U}(m)}^{-1}\right) _{l_{1}}^{k_{1}}*\left( W_{%
{\cal U}(m)}^{-1}\right) _{l_{2}}^{k_{2}}*...*\left( W_{{\cal U}%
(m)}^{-1}\right) _{l_{m}}^{k_{m}}\ .  \label{8a}
\end{eqnarray}

\subparagraph{Noncommutative Standard Model}

Let us go back now to the model of ref. \cite{Chaichian:2001py}.
Recall that
the 'minimal' gauge group for the NC Standard Model is $G_{NCSM}={\cal U}%
(3)\times {\cal U}(2)\times {\cal U}(1)$. We slightly modify the matter
content, however. Usual quarks and leptons are sitting in the following $%
G_{NCSM}$ multiplets,
\begin{eqnarray}
L &=&\left(
\begin{array}{c}
\nu  \\
e^{-}
\end{array}
\right) _{L}\sim (1,2,0);\ \ \ E=e_{L}^{c}\sim (1,1,-1);  \nonumber \\
Q &=&\left(
\begin{array}{c}
u \\
d
\end{array}
\right) _{L}\sim (3,\overline{2},0);\ \ \ U=u_{L}^{c}\sim (\overline{3}%
,1,+1);\ \ \ D=d_{L}^{c}\sim (\overline{3},1,0);  \label{9}
\end{eqnarray}
and similarly for the remaining generations. The operator of the
ordinary weak
hypercharge is a superposition of trace-$U(1)$ generators of $G_{NCSM}:$%
\[
Y=-\frac{2}{3}T^{0}_{{\cal U}(3)}-T^{0}_{{\cal U}(2)}-2T^{0}_{{\cal
U}(1)}\ .
\]
It is easy to see that $Y$ correctly reproduces the hypercharges (and hence
the electric charges) of ordinary quarks and leptons when acting on (\ref{9}%
). The above fermionic content is anomalous, however. To cancel
the anomalies it is sufficient to introduce a pair of ${\cal
U}(2)$-doublet lepton fields per generation:
\begin{equation}
L^{\prime }=\left(
\begin{array}{c}
E^{+} \\
N^{\prime }
\end{array}
\right) _{L}\sim (1,2,-1)\mbox{ and }L^{\prime \prime }=\left(
\begin{array}{c}
N^{\prime \prime } \\
E^{-}
\end{array}
\right) _{L}\sim (1,2,0).  \label{10}
\end{equation}
Remarkably, they are vector-like under the $G_{SM}$ subgroup of $G_{NCSM}$.
That means that upon the $G_{NCSM}$ symmetry breaking down to $G_{SM}$ these
extra states might pick up the masses and decouple from the low energy
spectrum. The relevant Yukawa interactions can be written using the Wilson
lines again:
\begin{equation}
\left( W_{{\cal U}(2)}*L^{\prime }*W_{{\cal U}(1)}^{-1}\right) ^{T}*\left(
W_{{\cal U}(2)}*L^{\prime \prime }\right) *\Phi _{{\cal U}(2)\times{\cal U}%
(1)}+h.c.  \label{11}
\end{equation}
where $\Phi _{{\cal U}(2)\times {\cal U}(1)}$ is ${\cal U}(2)\times {\cal U}%
(1)$ composite Higgsac field analogous of (\ref{8a}):
\begin{equation}
\Phi _{{\cal U}(2)\times {\cal U}(1)}=\frac{1}{2}\epsilon ^{j_{1}j_{2}}W_{{\cal U}%
(1)}*\phi _{[i_{1}i_{2}]}*\left( W_{{\cal U}(2)}^{-1}\right)
_{j_{1}}^{i_{1}}*\left( W_{{\cal U}(2)}^{-1}\right) _{j_{2}}^{i_{2}},
\label{12}
\end{equation}
and the proper contraction of gauge indices is understood. The vacuum
expectation value of this field, $<\Phi _{{\cal U}(2)\times {\cal U}(1)}>$,
provides spontaneous symmetry breaking: ${\cal U}(2)\times {\cal U}%
(1)\rightarrow SU(2)\times U(1)_{1-2},$ where the surviving
$U(1)_{1-2}$ is a linear combination of trace-$U(1)$ of ${\cal
U}(2)$ and ${\cal U}(1)$. At the same time the pair of left-handed
leptons acquires Dirac mass of the order of $<\Phi _{{\cal
U}(2)\times {\cal U}(1)}>$. To break $G_{NCSM}$ down
to $G_{SM}$ fully we must introduce one more Higgsac field, either $\Phi _{{\cal U}%
(3)\times {\cal U}(1)}$ or $\Phi _{{\cal U}(3)\times {\cal
U}(2)}$. Upon the condensation of these fields the only $U(1)$
that remains massless is the usual weak hypercharge field.

\subparagraph{Discussion and conclusions.}

We have proposed a mechanism for the spontaneous reduction of the
noncommutative gauge symmetry, i.e. ${\cal U}(n)\rightarrow
SU(n)$. This has been achieved through the
condensation of composite Higgsac fields (\ref{4}, \ref{8a}) in the trace-$%
U(1)$ noninvariant but $SU(n)$ preserving direction. An essential part of
our construction was the half-infinite Wilson lines that provide
gauge-invariant completion of the Higgsac mechanism proposed earlier in \cite
{Chaichian:2001py}.

The proposed mechanism offers new perspectives in realistic model building
based on the Weyl-Moyal NC gauge theories. In particular we have briefly
discussed NC Standard Model. Beside the spontaneous reduction of $G_{NCSM}=%
{\cal U}(3)\times {\cal U}(2)\times {\cal U}(1)$ down to
$G_{SM}=SU(3)\times SU(2)\times U(1)$ with $U(1)$ being the usual
weak hypercharge, we have demonstrated that anomalies can be
cancelled by the introduction of lepton pairs (per generation of
ordinary quarks and leptons) which are vector-like under the
$G_{SM}$ (not $G_{NCSM}$) gauge group. Moreover, the same Higgsac
field that provides the breaking $G_{NCSM}\rightarrow G_{SM}$,
couples in a gauge-invariant way to the extra lepton pairs and
provides their masses. Thus the low energy theory can be fully
reduced to the Standard Model with usual spectrum of ordinary
quarks and leptons.

The supersymmetric version of the NC Standard Model proposed in this
paper is a subject of its own interest which would remove
automatically the IR quadratic divergences arising from the UV/IR
mixing \cite{Minwalla:1999px, Matusis:2000jf}.

One can also construct grand unified models where certain features of the NC
Standard Model discussed here come out naturally. One is the NC
trinification model based on the gauge group ${\cal U}(3)\times {\cal U}%
(3)\times {\cal U}(3)$. Remarkably the standard minimal fermionic content of
the commutative trinification \cite{Glashow:1984gc} is automatically
anomaly-free in noncommutative case as well. For the spontaneous reduction $%
{\cal U}(3)\times {\cal U}(3)\times {\cal U}(3)\rightarrow $
$SU(3)\times SU(3)\times SU(3)$ one can use the gauge-invariant
Higgsac mechanism we have proposed in this paper.

\subparagraph{Acknowledgments.}

We are indebted to P. Pre\v{s}najder and M. M. Sheikh-Jabbari for
useful comments.

This work was supported by the Academy of Finland under the
projects 54023 and 104368. A.K. also acknowledges support from the
World Laboratory, Lausanne.

\end{document}